\documentclass[pra,a4paper,aps,twocolumn,showpacs]{revtex4}
\usepackage[dvips]{graphicx}
\usepackage{ae}
\usepackage[T1]{fontenc}
\usepackage[ansinew]{inputenc}
\usepackage{amsmath}
\usepackage{amssymb}
\usepackage{graphicx}
\usepackage{caption}
\usepackage{color}
\usepackage[colorlinks=true,linkcolor=red,citecolor=blue]{hyperref}
\usepackage{lscape}
\begin{document}
	\title{Eavesdropping a Quantum Key Distribution network using sequential quantum unsharp measurement attacks}
	
\author{Yash Wath}
\email{yashprashantwath@students.iisertirupati.ac.in}
\affiliation{Indian Institute of Science Education and Research, Tirupati 517507, India}

\author{Hariprasad M}
\email{hpmadathil@gmail.com}
\affiliation{Pondicherry University, Chinna Kalapet, Kalapet, Puducherry 605014, India}

\author{Freya Shah}
\email{fs1132429@gmail.com}
\affiliation{Udgam School for Children, Ahmedabad 380051, India}

\author{Shashank Gupta}
\email{shashankg687@bose.res.in}
\affiliation{S. N. Bose National Centre for Basic Sciences, Block JD, Sector III, Salt Lake, Kolkata 700 106, India}

\date{\today}

\begin{abstract}
We investigate the possibility of eavesdropping a quantum key distribution network by local sequential quantum unsharp measurement attacks by the eavesdropper. In particular, we consider a pure two-qubit state shared between two parties Alice and Bob, sharing quantum steerable correlations that form the one-sided device-independent quantum key distribution network. One qubit of the shared state is with Alice and the other one while going to the Bob's place is intercepted by multiple sequential eavesdropper who perform quantum unsharp measurement attacks thus gaining some positive key rate while preserving the quantum steerable correlations for the Bob. In this way, Bob will also have positive secret key rate although reduced. However, this reduction is not that sharp and can be perceived due to decoherence and imperfection of the measurement devices. At the end, we show that unbounded number of eavesdropper can also get secret information in some specific scenario.

\end{abstract}

\maketitle

\section{INTRODUCTION}

Quantum network is a network of observers that are connected by quantum channels. The main idea is to use quantum channels to transfer secret information from one observer to another. This forms the quantum key distribution (QKD) network \cite{Gisin2002}. There are two distinct classes of QKD protocols in the literature: 1. Prepare and measure \cite{BB84, Bennet92}, and 2. Entanglement-assisted \cite{ekert}. One distinct advantage that entanglement-assisted QKD protocols offers is the ability to check its security based on monogamy relations via the violation of steering or Bell inequalities.

Quantum key distribution protocols have been proven to be secure against eavesdropping and noise \cite{scarani2009}. The rigorous security proof was by Mayers \cite{Mayers97,Mayers2001}. The other development in this direction were based on entanglement distillation \cite{Deutsch96,Lo99}. The work by Shor and Preskill \cite{Shor2000} used the Mayer's result and the ideas of quantum error correction to prove the security of the BB84 protocol.

With the discovery of photon-number-splitting (PNS) attacks \cite{Brassard2000}, security proofs came to deal with the loophole of theoretical single-photon source \cite{Norbert2000}. With increasing quantum technological advancement, the possibility of newer kind of attacks by the adversary raises the question on the security of such QKD protocols time and again. Here, we investigate the security of one-sided device-independent quantum key distribution (1S-DIQKD) networks \cite{Branciard12} under the possibility of quantum unsharp measurement attacks by the Eavesdropper.

Quantum steering \cite{steerreview,Schr1,Schr2,Wise1,Wise2,Cavalcanti1,Cavalcanti2,SGupta21} is an intermediate correlation between entanglement and Bell-nonlocality, was discovered by E. Shr\"odinger as an interpretation of the famous work by Einstein-Podolsky-Rosen \cite{epr}. It is the key resource behind 1S-DIQKD whose security is based on the monogamy of the quantum steering \cite{Pramanik}. Recently, it has been shown that quantum steering can be shared by multiple observers using unsharp measurements \cite{Sasmal18,Gupta21}. This raises a critical question on the security of 1S-DIQKD network, what if the eavesdropper perform unsharp measurements and gain the secret information by still preserving the steerable correlations?

In the present work, we address the question raised above by finding the upper bound on the number of Eves that can share the secret key rate by finding the violation of fine-grained steering inequality \cite{Pramanik} such that the Bob has secret key rate value more than any Eve in the network. Since, the Bob is getting a positive key rate, he will not terminate the protocol and perceive the reduction in the key rate value as an effect of the noise or the imperfection of the measurement devices. We also present a strategy by which an unbounded number of Eves can have access to secret information in some specific scenario \cite{Curchod17}, thus raising the magnitude of the criticality of such QKD protocols.

The plan of the paper is as follows: in Section \ref{s2} we recapitulate and present the basic tools that are required for further investigation. The measurement framework involving multiple sequential Eves used in this paper is also described in this Section. In Section \ref{s3}, we present the main analysis of this paper, and discuss the results obtained in the context of sequential eavesdropping of the 1S-DIQKD network when the initially shared two-qubit state is maximally entangled. We also discuss a strategy that gives rise to unbounded number of eavesdroppers. Finally, we conclude in Section \ref{s4}.

\section{Preliminaries} \label{s2}

In this section, we discuss in brief the concept of EPR steering and  its connection with the one-sided device-independent quantum key distribution (1S-DIQKD). We then briefly discuss the quantum sequential measurement attacks in the context of eavesdropping.

\subsection{Steering and 1S-DIQKD}
 Consider that a bipartite state $\rho$ is shared among two observers, say, Alice and Bob. In the steering scenario, Alice tries to steer Bob's particles via her input and output measurement choices. Alice's measurement operators are denoted by $A_{a|x}$, where $x$ is the choice of her input and $a$ is the outcome. After Alice's measurement, each element of the set $\{\sigma_{a|x}^{B}\}_{a,x}$ of unnormalized conditional states on Bob end is given by,
 \begin{equation}
	\sigma_{a|x}^{B}= \text{Tr}_A\big[(A_{a|x} \otimes \mathbf{I}_B ) \, \rho \big].
\label{assemblage1}
\end{equation}

This set of conditional states is called an assemblage. The entangled or the separable nature of the state imposes constraint on the observed assemblage. When the initial state $\rho$ is not entangled, then  each element (\ref{assemblage1}) of the assemblage $\{\sigma_{a|x}^{B}\}_{a,x}$ is of the following form,
\begin{equation}
	\sigma_{a|x}^{B} = \sum_{\lambda}p_{\lambda}p_{\lambda}(a|x)\rho_{\lambda}^B \,  \quad \forall \, a,x. 
				\label{ass1}
\end{equation}
Here, $p_{\lambda}(a|x)$ denotes the probability of getting the outcome $a$ when Alice performs the measurement denoted by $x$ on the local state $\rho_{\lambda}^A$. If any element of the assemblage $\{\sigma_{a|x}^{B}\}_{a,x}$ cannot be written in the form (\ref{ass1}), then the assemblage indicates that the state $\rho$ possesses entanglement. In other words, when an element  of the assemblage $\{\sigma_{a|x}^{B}\}_{a,x}$ cannot be written in the form (\ref{ass1}), then the assemblage demonstrate EPR steering from Alice to Bob.

In literature, steering has been detected using logical contradiction, uncertainty relations and inequalities. Here, we use the optimal fine-grained steering criteria to demonstrate steering and and use the monogamy relation to study the 1S-DIQKD of steerable states \cite{Pramanik}. Alice performs a dichotomic measurements $\{A_{a|x}\}$ s.t. $a, x \in \{0, 1 \}$ and communicates her input choice and outcome to Bob. In order to check whether the state is steerable,Bob performs measurements in two mutually unbiased basis, say, $\{\sigma_x, \sigma_z\}$ with outcome $b \in \{0, 1\}$. The shared state $\rho$ is steerable, if the conditional probability distribution violates the following relation:

\begin{equation}
\frac{1}{2} \big[ P(B_{b|0}|A_{a|0}) + P(B_{b|1}|A_{a|1}) \big] \leq \frac{3}{4}
\label{fgi}
\end{equation}

Now, if the shared state $\rho$ between Alice and Bob is maximally steerable, then none of these systems can be steerable with any other system \cite{Pramanik}. This nature is called monogamy of steerable states. This behaviour of steerable states put a lower bound on the secret key rate in 1S-DIQKD (one-sided device independent because one of the systems is not trusted, here Alice). The lower bound on the secret key rate (r), corresponding to the shared state $\rho$ which violates the above inequality (\ref{fgi}) is given by \cite{Pramanik}.

\begin{equation}
r \geq \text{log}_2\Bigg[\frac{\frac{3}{4}+\delta}{\frac{3}{4}-\delta}\Bigg]
\label{sk}
\end{equation}
Here, $\delta$ ($\in (0, \frac{1}{4}])$ is the degree of violation of the inequality (\ref{fgi}).

\subsection{Sequential quantum unsharp measurement attacks} \label{scenario}
We now describe the idea of quantum unsharp measurement attacks performed by the eavesdropper for sequential eavesdropping the 1S-DIQKD network.



 Consider  a bipartite system of state $\rho$ consisting of spatially separated two spin-$\frac{1}{2}$ particles.  In this case, we consider multiple Eves (e$^1$, e$^2$, $\cdots$, e$^n$) perform measurements on the second particle sequentially. On the other hand, a single Alice perform projective measurements on the first particle. Here we ask the following question:

$\bullet$ \textit{How may Eves can gain positive secret key rate while preserving the quantum steerable correlations (positive key rate greater than any of the Eves) for the Bob?}

Since our aim is to explore how many Eves can gain positive secret key rate through the violation of Fine-grained steering inequality (\ref{fgi}), multiple Eves cannot perform projective measurements. If any Eve performs a projective measurement, then the EPR steerability of the state will be completely lost and Bob will know that there is an eavesdropping in the network. Hence, each $n$ Eves in the sequence should perform weak measurements.  



\begin{figure}[t!]
\centering
\includegraphics[scale=0.55]{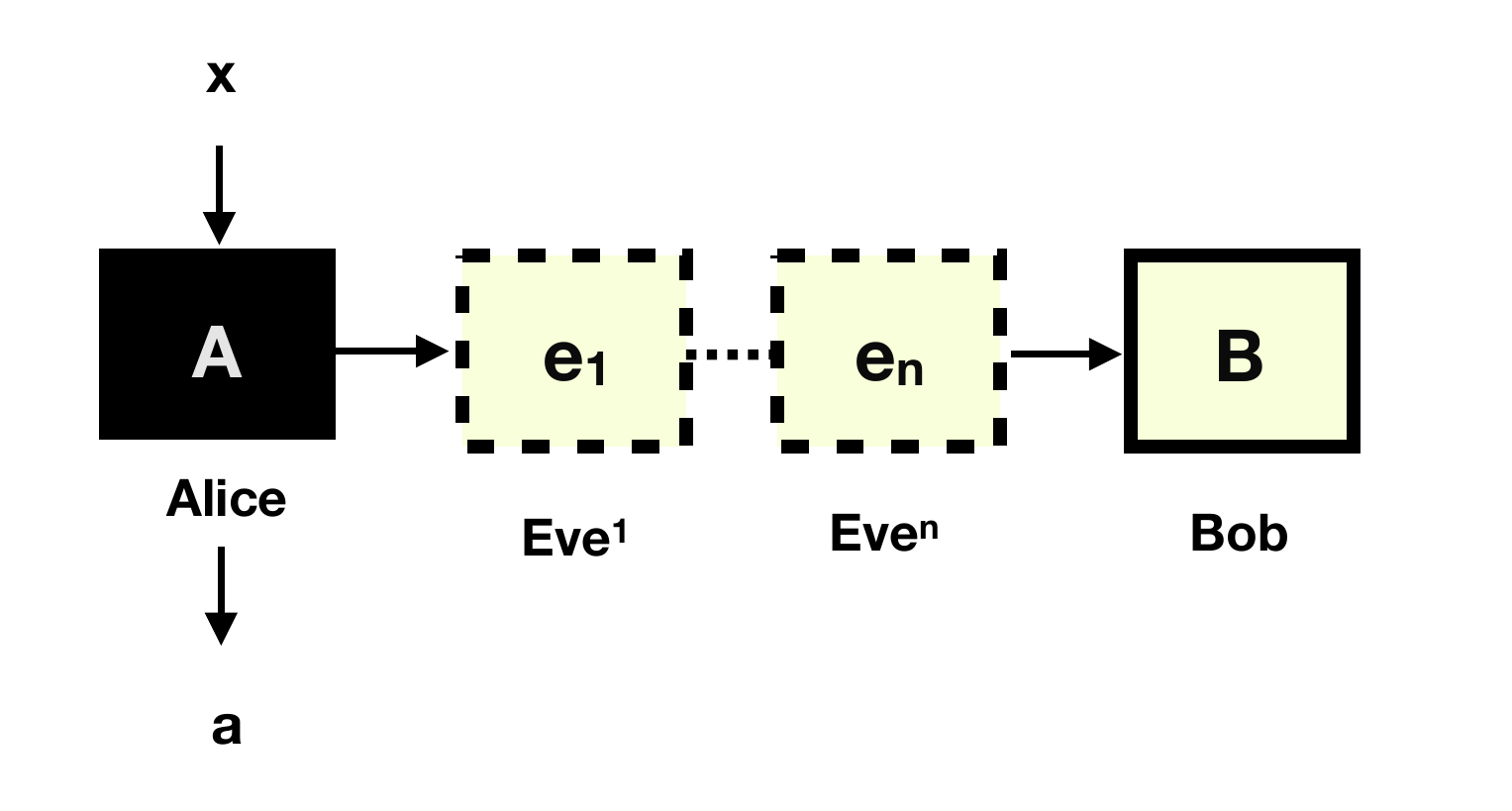}
\caption{(Color Online) Sequential eavesdropping of 1S-DIQKD network by multiple Eves. Two spatially separated spin-$\frac{1}{2}$ particles,  prepared in the two qubit state $\rho$,  are shared between Alice and Bob but secretly tapped by multiple Eves (Eve$^1$, Eve$^2$, $\cdots$, Eve$^n$). }
\label{fig1}
\end{figure}

The aforementioned scenarios are depicted in Figure \ref{fig1}. 

Next, let us briefly discuss the quantum unsharp measurement attacks used by the eavesdroppers (For details, see \cite{sygp,majumdar,Sasmal18}). In a sharp projective measurement attack, one obtains the maximum amount of information at the cost of maximum disturbance to the state which is detected by the Alice and Bob. On the other hand, in our scenario, Eve$^m$  passes on the respective particle to Eve$^{m+1}$ after getting some positive key rate. Hence, in this case, Eve$^m$ needs to access the positive secret key rate by disturbing the state minimally so that Eve$^{m+1}$ (or, Bob at the end) can also get positive secret key rate. This can be achieved by unsharp measurements \cite{sygp} which may be characterized by two real parameters: the quality factor $(F)$ and the precision $(G)$ of the measurement. $F$ quantifies the extend to which the initial state of the system remains undisturbed during  the measurement process and $G$ quantifies the information gain due to the measurement. In case of projective measurement, $F =0$, $G=1$. For dichotomic measurements on a qubit system, the optimal trade-off relation is given by, $F^2 + G^2 =1$ \cite{sygp}. 

The above optimal trade-off relation between information gain and quality factor is achieved under unsharp measurement formalism \cite{majumdar,Sasmal18}. Unsharp measurement \cite{pb1,pb2} is one particular class of  positive operator valued measurements (POVM) \cite{pb1,pb2}. 
Consider a dichotomic observable $O$ $=$ $\Pi_{0} - \Pi_{1}$ with outcomes $0$ and $1$, where $\Pi_{0}$ and $\Pi_{1}$ denote the projectors associated with the outcomes $0$ and $1$ respectively, with $\Pi_{0} + \Pi_{1} = \mathbb{I}$ and $\Pi_{0/1}^2 = \Pi_{0/1}$. Given the observable $O$, a dichotomic unsharp observable  $O^{\lambda} = E^{\lambda}_0 - E^{\lambda}_1$ \cite{umn1,umn2} can be defined which is associated with the sharpness parameter $\lambda \in (0, 1]$. Here, the effect operators $E^{\lambda}_{0/1}$ are given by, 
\begin{equation}
E^\lambda_{0/1} = \lambda \, \Pi_{0/1} + (1-\lambda) \frac{\mathbb{I}_2}{2}.
\end{equation} 
This is obtained by mixing projective measurements with white noise. The probability of getting the outcomes $0$ and $1$, when the above unsharp measurement is performed on the state $\rho$, are given by $\text{Tr}[\rho E^\lambda_{0}]$ and $\text{Tr}[\rho E^\lambda_{1}]$ respectively. Using the generalized von Neumann-L\"{u}ders transformation rule \cite{pb1}, the states after the measurements, when the outcomes $0$ and $1$ occurs, are given by, $\dfrac{\sqrt{E^\lambda_{0}} \rho \sqrt{E^\lambda_{0}}}{\text{Tr}[E^\lambda_{0} \rho]}$ and $\dfrac{\sqrt{E^\lambda_{1}} \rho \sqrt{E^\lambda_{1}}}{\text{Tr}[E^\lambda_{1} \rho]}$ respectively.

Unsharp measurements has already been demonstrated using trapped-ion \cite{une1} and photonic systems \cite{exp1,exp2,Choi2020,exp4,rac1}. Here, using an appropriate interferometer one can realize the unsharp measurement formalism. Further, the sharpness parameter can be controlled by fine-tuning the arrangement of various components in the interferometer. This raises a critical question related to the security of QKD networks where multiple Eves in the sequence perform unsharp measurement attacks where as Alice and Bob perform sharp measurements..

\section{Eavesdropping by multiple sequential Eves}\label{s3}

Using the formalism discussed in the previous section, we are now in a position to explore the maximum number of Eves that can gain the positive secret key rate such that Bob also gets the positive key rate greater than any of the Eves. Before proceeding, let us recapitulate some previous relevant results on sequential detection of steering. 

It has been shown \cite{Sasmal18} that at most two Bob can sequentially be steered with a single Alice when the singlet state is initially shared and each party is performing dichotomic measurements. Now, in order to demonstrate steering, entanglement is necessary and hence, it can be concluded that two Bob can sequentially detect entanglement of the singlet state in a one-sided DI scenario with a single Alice. These results are probed through  the quantum violation of the analogous CHSH inequality \cite{Cavalcanti15}. 

Here, we ask a similar question, but in the one-sided device-independent quantum key distribution scenario. Let Alice perform dichotomic projective measurement of the spin component observable  in the direction $\hat{x}_0$, or  $\hat{x}_1$. Bob performs dichotomic projective measurement of the spin component observable  in the direction $\hat{y}_0$, or $\hat{y}_1$.  Eve$^m$ (where $m$ $\in \{1, 2, \cdots, n\}$) performs dichotomic unsharp measurement of the spin component observable  in the direction $\hat{z}_0^m$, or $\hat{z}_1^m$. The outcomes of each measurement is $0$ $ \text{or}$  $1$.

The projectors associated with Alice's sharp spin component measurement in the direction $\hat{x}_i$ (with $i$ $\in$ $\{0,1\}$) are given by, $A_{a|\hat{x}_i} = \dfrac{\mathbb{I}_2+ (-1)^a \, \hat{x}_i \cdot \vec{\sigma}}{2}$ (with $a$ $\in$ $\{0, 1\}$ being the outcome of Alice's sharp measurement). Similarly, the projectors associated with Bob's sharp spin component measurement in the direction $\hat{y}_j$ (with $j$ $\in$ $\{0,1\}$) can be written as $B_{b|\hat{y}_j} = \dfrac{\mathbb{I}_2+ (-1)^b \, \hat{y}_j \cdot \vec{\sigma}}{2}$ (with $b$ $\in$ $\{0, 1\}$ being the outcome of Bob's sharp measurement). Here $\vec{\sigma}$ = $(\sigma_1, \sigma_2, \sigma_3)$ is a vector composed of three
Pauli matrices. The directions $\hat{x}_i$ and $\hat{y}_j$ can be expressed as,
\begin{equation}
\label{alicedir}
\hat{x}_i = \sin \theta^{x}_i \cos \phi^{x}_i \hat{X} + \sin \theta^{x}_i \sin \phi^{x}_i \hat{Y} + \cos \theta^{x}_i \hat{Z},
\end{equation}
and 
\begin{equation}
\label{bobdir}
\hat{y}_j = \sin \theta^{y}_j \cos \phi^{y}_j \hat{X} + \sin \theta^{y}_j \sin \phi^{y}_j \hat{Y} + \cos \theta^{y}_j \hat{Z},
\end{equation}
where $i, j \in \{0, 1\}$; $0 \leq \theta^{x}_i  \leq \pi$; $0 \leq \phi^{x}_i  \leq 2 \pi$; $0 \leq \theta^{y}_j  \leq \pi$; $0 \leq \phi^{y}_j  \leq 2 \pi$. $\hat{X}$, $\hat{Y}$, $\hat{Z}$ are three orthogonal unit vectors in Cartesian coordinates. 

The effect operators associated with Eve$^m$'s ($m$ $\in$ $\{1, 2, \cdots, n \}$) unsharp measurement of spin component observable in the direction $\hat{z}^m_k$ (with $k$ $\in$ $\{0,1\}$) are given by,
\begin{equation}
E^{\lambda_m}_{c^m|\hat{z}^m_k} = \lambda_{m}\frac{\mathbb{I}_2+ (-1)^{c^{m}} \hat{z}^m_k \cdot \vec{\sigma}}{2}+(1-\lambda_{m})\frac{\mathbb{I}_2}{2},
\end{equation}
with $c^m$ $\in$ $\{0, 1\}$ being the outcome of Eve$^m$'s unsharp measurement and $\lambda_m$ ($0 < \lambda_m \leq 1$) is the sharpness parameter corresponding to Eve$^m$'s unsharp measurement. The direction $\hat{z}^m_k$ is given by,
\begin{equation}
\label{evemdir}
\hat{z}^m_k = \sin \theta^{z^m}_k \cos \phi^{z^m}_k \hat{X} + \sin \theta^{z^m}_k \sin \phi^{z^m}_k \hat{Y} + \cos \theta^{z^m}_k \hat{Z},
\end{equation}
where $k \in \{0, 1\}$; $0 \leq \theta^{z^m}_k  \leq \pi$; $0 \leq \phi^{z^m}_k  \leq 2 \pi$.

There are conditional probabilities appearing in the inequality (\ref{fgi}). In the following, we compute these conditional probabilities between Eve$^m$ and Alice. 

The conditional probability of occurrence of the outcome  $c^1$ when Eve$^1$ performs unsharp measurement of spin component observable along the direction $\hat{z}^1_1$ given that Alice performs projective measurements of spin component observables along the directions $\hat{x}_1$ and got the outcome $a$ on the shared bipartite state $\rho$, is given by,
\begin{equation}
P(E^{1}_{c^1|z^1_k}|A_{a|x_i}) =\frac{\text{Tr}\Bigg[\Bigg\{ A_{a|\hat{x}_i} \otimes E^{\lambda_1}_{c^1|\hat{z}^1_k}  \Bigg\}  \cdot \rho \Bigg]}{\text{Tr}\Bigg[\Bigg\{ A_{a|\hat{x}_i} \otimes \mathbb{I}_2  \Bigg\}  \cdot \rho \Bigg]}.
\end{equation}

After performing unsharp measurement, Eve$^1$ passes her particle to Eve$^2$. The unnormalized post-measurement reduced state at Eve$^2$'s end, after Eve$^1$ gets the outcome $c^1$ by performing unsharp measurement of spin component observable along the direction $\hat{z}^1_1$ and Alice get the outcomes $a$  by performing sharp measurements of spin component observables along the directions $\hat{x}_1$  is given by,
\begin{align}
\rho_{un}^{e^2} =& \text{Tr}_{A} \Bigg[ \Bigg\{\frac{\mathbb{I}_2 + (-1)^a \hat{x}_i \cdot \vec{\sigma}}{2} \otimes \sqrt{E^{\lambda_1}_{c^1|\hat{z}^1_k}}  \Big\}  \nonumber \\ 
& \cdot \rho \cdot \Bigg\{\frac{\mathbb{I}_2 + (-1)^a \hat{x}_i \cdot \vec{\sigma}}{2} \otimes \sqrt{E^{\lambda_1}_{c^1|\hat{z}^1_k}}  \Big\} \Bigg],
\end{align}
where,
\begin{align}
\sqrt{E^{\lambda_1}_{c^1|\hat{z}^1_k}} &= \sqrt{\dfrac{1+\lambda_1}{2}} \Bigg( \dfrac{\mathbb{I}_2 + (-1)^{c^1} \hat{z}^1_k \cdot \vec{\sigma}}{2} \Bigg) \nonumber \\
& +\sqrt{\dfrac{1- \lambda_1}{2}} \Bigg( \dfrac{\mathbb{I}_2 - (-1)^{c^1} \hat{z}^1_k \cdot \vec{\sigma}}{2} \Bigg).
\end{align}
In order to get the reduced state, the partial trace has been taken over the Alice's subsystem.

Now, Eve$^2$ also performs unsharp measurement attack (associated with sharpness parameter $\lambda_{2}$) of spin component observable along the direction $\hat{z}^2_k$ on the reduced state $\rho_{un}^{e^2}$ and gets the outcome $c^2$. The joint probability of occurrence of the outcomes $a$, $c^1$, $c^2$, when Eve$^1$, Eve$^2$ perform unsharp measurements of spin component observables along the directions $\hat{z}^1_k$, $\hat{z}^2_k$ respectively and Alice perform projective measurements of spin component observables along the directions $\hat{x}_i$ , is given by,
\begin{equation}
P(A_{a|x_i}, E^1_{c^1|z^1_k}, E^2_{c^2|z^2_k}) = \text{Tr} \Big[ E^{\lambda_2}_{c^2|\hat{z}^2_k} \cdot \rho^{e^2}_{un} \Big].
\end{equation}
 From this expression, the joint probability of obtaining the outcomes $c^2$, $a$, by Eve$^2$, Alice, respectively, can be calculated as,
\begin{align}
& P(A_{a|x_i}, E^2_{c^2|z^2_k}) \nonumber \\
&= \sum_{c^1 = 0}^{ 1} \Big[ P(A_{a|x_i}, E^1_{c^1|z^1_1}, E^2_{c^2|z^2_k})P(\hat{z}^1_1)\nonumber \\
&+P(A_{a|x_i}, E^1_{c^1|z^1_2}, E^2_{c^2|z^2_k})P(\hat{z}^1_2) \Big].
\label{avprob2}
\end{align}
Here, $P(\hat{z}^1_k)$ is the probability of Eve$^1$'s unsharp measurement of spin component observable in the direction $\hat{z}^1_k$ ($k \in \{0, 1\}$). Since, we restrict ourselves to unbiased input scenario, the two measurement settings for Eve$^1$ are equally probable, i.e.,  $P(\hat{z}^1_1)$ = $P(\hat{z}^1_2)$  = $\frac{1}{2}$. Now, the conditional probability of occurrence of the outcomes $c^2$ when Eve$^2$ perform unsharp measurements of spin component observable along the direction $\hat{z}^2_k$ given that Alice perform projective measurements of spin component observables along the directions $\hat{x}_i$ and got the outcome $a$, is given by,

\begin{equation}
P(E^{2}_{c^2|z^2_k}|A_{a|x_i}) =\frac{P(A_{a|x_i}, E^2_{c^2|z^2_k})}{\text{Tr}\Bigg[\Bigg\{ A_{a|\hat{x}_i} \otimes \mathbb{I}_2  \Bigg\}  \cdot \rho \Bigg]}.
\label{avcond2}
\end{equation}

Using the expression (\ref{avcond2}) for the conditional probability, the two terms appearing in inequality (\ref{fgi}) can be easily calculated for detection of EPR steering by Eve$^2$. Using the violation of the inequality one can easily determine the positive secret key rate value for Eve$^2$ in terms of the sharpness parameter $\lambda_2$. Following the above-mentioned approach, the two terms appearing on the left hand side of the inequality (\ref{fgi}) in the context of  Eve$^m$, Alice and Bob can be calculated. In the following, we consider that the initial state is a two-qubit maximally entangled state.

\subsubsection{When the two-qubit maximally entangled state is initially shared} \label{sub1}
Let us consider that the two-qubit maximally entangled state given by $\rho_{\text{s}} = | \psi_{\text{s}} \rangle \langle \psi_{\text{s}} |$  is shared between Alice and Bob, where
\begin{equation}
|\psi_{\text{s}} \rangle = \frac{1}{\sqrt{2}} ( |00 \rangle + | 11 \rangle ).
\label{singlet}
\end{equation}
Here, $|0\rangle$ and $|1\rangle$ denotes two mutually orthonormal states in $\mathbb{C}^2$. Here, multiple Eves perform sequential unsharp measurement attacks on the second qubit.

 At first, we will find out the maximum number of Eves, who can access positive secret key rate such that the Bob will also have positive secret key rate greater than any of the Eves. This will be probed through the quantum violation of the inequality (\ref{fgi}) and the relation (\ref{sk}).


 We start by finding out  whether Eve$^1$ and Eve$^2$ can sequentially access positive secret key rate. In other words, we will find out whether Eve$^1$ and Eve$^2$ can sequentially have the secret key rate (\ref{sk}) value say `r' such that Bob has the secret key rate $\geq r$. In this case, the measurements of the Eves' are unsharp where as that of Alice and Bob are sharp. We observe that, for example, when  Eve$^1$ and Eve$^2$ get $r = 0.1$,  then Bob gets $r = 0.634$. This happens for the following choices of measurement settings by Alice:  $( \theta^{x}_1$, $\phi^{x}_1$, $\theta^{x}_2$, $\phi^{x}_2$ ) $\equiv$ $( 0$, $0$, $\frac{\pi}{2}$, $0$ ), with $\lambda_1 = 0.552$ and $\lambda_2 =0.602 $. Note that, here the choices of measurement settings by Eves and Bob are in two mutually unbiased basis ($\sigma_x$ and $\sigma_z$). Hence, we can conclude that Eve$^1$ and Eve$^2$ can access secret key rate value of 0.1 such that Bob also have positive value of secret key rate. Note that the secret key rate at Bob's end would be higher if there were no Eve's. However, this reduction in the value of secret key rate is not sharp and Bob may perceive it due to decoherence or imperfection of the measurement devices.

Next, we ask whether Eve$^1$, Eve$^2$ and Eve$^3$ can sequentially access positive secret key rate such that Bob also have positive secret key rate. We observe that, when Eve$^1$, Eve$^2$ and Eve$^3$ get $r = 0.2$, then Bob gets $r = 0.269$.  This happens for the same choices of measurement settings by Alice as mentioned above, with $\lambda_1 = 0.604$, $\lambda_2 = 0.672$ and $\lambda_3 = 0.772$. Hence, Eve$^1$, Eve$^2$ and Eve$^3$ can access secret key rate when the maximally entangled two-qubit state is initially shared.

Proceeding this way, we find the upper bound on the number of Eves' that can access secret key rate of value 0.1, 0.2 and 0.3 respectively such that the Bob has secret key rate of value more than that each sequential Eve can access. These results are summarized in Table \ref{tab1}, \ref{tab2} and \ref{tab3}. The permissible range of each $\lambda_m$ depends on the values $\lambda_1$, $\lambda_2$, $\cdots$, $\lambda_{m-1}$. In the table, we have presented the permissible range of each $\lambda_m$ for the minimum permissible value of each $\lambda_1$, $\lambda_2$, $\cdots$, $\lambda_{m-1}$. The permissible range of $\lambda_m$ will be smaller than this if we take other value $\lambda_i$ > $\lambda_i^{\text{min}}$ $\forall$ $i < m$, and the maximum number of Eves may get reduced. Note that as the value of the secret key rate associated with Eves increases the number of Eves that can perform these quantum unsharp measurement attacks decreases as even higher value of the key rate is demanded at the Bob's end. Hence, Bob can assure the security of the QKD network by having higher secret key rate and terminating the protocol in adverse scenario.

{\centering
	\begin{table}[t]\footnotesize
			\begin{tabular}{|c | c |} 
				\hline 
				& \textbf{Secret key rate for each Eve is 0.1}  \\
				\hline
				{\bf Eve$^m$} & {\bf Permissible $\lambda_{m}$ range}    \\ [0.5ex] 
				\hline
				\hline
				Eve$^1$ &  1 $\geq \lambda_1 > \lambda_1^{\text{min}}$ = 0.552   \\ 
				\hline
				Eve$^2$ &1 $\geq \lambda_2 > \lambda_2^{\text{min}}$ = 0.602   \\
				& when $\lambda_i = \lambda_i^{\text{min}}$ $\forall$ $i < 2$ \\
				\hline
				Eve$^3$ &1 $\geq \lambda_3 > \lambda_3^{\text{min}}$ = 0.67    \\
				& when $\lambda_i = \lambda_i^{\text{min}}$ $\forall$ $i < 3$  \\
				\hline
				Eve$^4$ &1 $\geq \lambda_4 > \lambda_4^{\text{min}}$ = 0.768    \\
				& when $\lambda_i = \lambda_i^{\text{min}}$ $\forall$ $i < 4$  \\
				\hline
				Eve$^5$ & No valid range for $\lambda_5$   \\
				&  for any $\lambda_i$ with $i<5$ \\ [1ex]
				\hline
			\end{tabular}
			\caption{The permissible ranges of the sharpness parameters $\lambda_m$ (where $0 < \lambda_m \leq 1$) of Eve$^m$ in order to access secret key rate $r = 0.1$ when the two-qubit maximally entangled state is initially shared such that the Bob has a positive secret key rate $r = 0.172$ $(> 0.1)$ at the end.}
			\label{tab1}
	\end{table}
}

{\centering
	\begin{table}[t]\footnotesize
			\begin{tabular}{|c | c |} 
				\hline 
				& \textbf{Secret key rate for each Eve is 0.2}  \\
				\hline
				{\bf Eve$^m$} & {\bf Permissible $\lambda_{m}$ range}    \\ [0.5ex] 
				\hline
				\hline
				Eve$^1$ &  1 $\geq \lambda_1 > \lambda_1^{\text{min}}$ = 0.604   \\ 
				\hline
				Eve$^2$ &1 $\geq \lambda_2 > \lambda_2^{\text{min}}$ = 0.672   \\
				& when $\lambda_i = \lambda_i^{\text{min}}$ $\forall$ $i < 2$ \\
				\hline
				Eve$^3$ &1 $\geq \lambda_3 > \lambda_3^{\text{min}}$ = 0.772    \\
				& when $\lambda_i = \lambda_i^{\text{min}}$ $\forall$ $i < 3$  \\
				\hline
				Eve$^4$ & No valid range for $\lambda_4$   \\
				&  for any $\lambda_i$ with $i<4$ \\ [1ex]
				\hline
			\end{tabular}
			\caption{The permissible ranges of the sharpness parameters $\lambda_m$ (where $0 < \lambda_m \leq 1$) of Eve$^m$ in order to access secret key rate $r = 0.2$ when the two-qubit maximally entangled state is initially shared such that the Bob has a positive secret key rate $r = 0.269$ $(> 0.2)$ at the end.}
			\label{tab2}
	\end{table}
}

{\centering
	\begin{table}[t]\footnotesize
			\begin{tabular}{|c | c |} 
				\hline 
				& \textbf{Secret key rate for each Eve is 0.3}  \\
				\hline
				{\bf Eve$^m$} & {\bf Permissible $\lambda_{m}$ range}    \\ [0.5ex] 
				\hline
				\hline
				Eve$^1$ &  1 $\geq \lambda_1 > \lambda_1^{\text{min}}$ = 0.655   \\ 
				\hline
				Eve$^2$ &1 $\geq \lambda_2 > \lambda_2^{\text{min}}$ = 0.747   \\
				& when $\lambda_i = \lambda_i^{\text{min}}$ $\forall$ $i < 2$ \\
				\hline
				Eve$^3$ & No valid range for $\lambda_3$   \\
				&  for any $\lambda_i$ with $i<3$ \\ [1ex]
				\hline
			\end{tabular}
			\caption{The permissible ranges of the sharpness parameters $\lambda_m$ (where $0 < \lambda_m \leq 1$) of Eve$^m$ in order to access secret key rate $r = 0.3$ when the two-qubit maximally entangled state is initially shared such that the Bob has a positive secret key rate $r = 0.447$ $(> 0.3)$ at the end.}
			\label{tab3}
	\end{table}
}

 Next, we devise a strategy where unbounded number of Eves can access the positive secret key rate when the initial shared state is a non-maximally entangled two-qubit pure state. 

\subsubsection{Eavesdropping by unbounded number of Eves}
Here, let us consider that the two-qubit non-maximally entangled state $\rho_{\theta} = | \psi_{\theta} \rangle \langle \psi_{\theta} |$ is initially shared between Alice and Bob, where
\begin{equation}
| \psi_{\theta} \rangle = \cos(\theta) | 00\rangle + \sin(\theta) |11\rangle.
\label{nmstate}
\end{equation} 

The fine-grained inequality (\ref{fgi}) has following properties: \textbf{1.} Any non-maximally entangled pure state of the form (\ref{nmstate}) maximally violates the inequality for the following measurements by the two parties:
\begin{align}
A_1 &= \sigma_z, \hspace{3.3 cm} B_1 = \sigma_z \nonumber \\
A_2 &= \cos(2\theta) \sigma_z + \sin(2\theta) \sigma_x, \quad B_2 = \sigma_x
\end{align} 
\textbf{2.} When maximally violated, the value of the secret key rate is also maximum i.e. 1. However, in general we may not get maximal violation. In that case, one can show that the secret key rate is a continuous function of the value of the inequality close to the maximum violation.

\textit{Strategy for unbounded eavesdropping:} Each Eve chooses between measurements of $\sigma_z$ and $\sigma_x^{\lambda_m}$ with outcomes $c^k \in \{0, 1\}$ ($\sigma_x^{\lambda_m}$ with 0 outcome is $M_{0}^{\dagger} M_{0}$, where $M_{0} = \cos(\lambda_m) |0\rangle \langle 0| + \sin(\lambda_m) |1\rangle \langle 1|$. 0 and 1 are the eigenbasis of the $\sigma_x$ operator). The parameters $\lambda_m$ are fixed before the beginning of the QKD protocol. The initial entangled state shared between Alice and Eve$^1$ is | $\psi^1_{\theta_1}\rangle$ [see Eq. (\ref{nmstate}) with $\theta = \theta_1$]. If the first unsharp measurement of the operator $\sigma_x^{\lambda_1}$ is made by Eve$^1$ on the initial state | $\psi^1_{\theta_1}\rangle$, the post-measured state is of the form:
\begin{equation}
	|\psi_{c^1}^2 (\theta_1, \lambda_1)\rangle = U_A^{c^1}(\theta_1, \lambda_1) \otimes V_E^{c^1}(\theta_1, \lambda_1) \Big(c|00\rangle + s|11\rangle\Big)
\end{equation}
where, c = $\cos(\theta_{c^1}(\theta_1, \lambda_1))$, s = $\sin(\theta_{c^1}(\theta_1, \lambda_1))$, the two unitaries $U_A^{c^1}$ and $V_E^{c^1}$ and the angle $\theta_{c^1}(\theta_1,\lambda_1) \in (0, \pi/4]$ depend on the outcome $c^1$ and the angles $\theta_1$ and $\lambda_1$.

After his measurement, Eve$^1$ applies the unitary ($V_E^{c^1})^{\dagger}$ conditioned on his outcome $c^1$, on the post-measured state going to Eve$^2$. This allows Eve$^2$ to use the same two measurements $\sigma_x^{\lambda_2}$ and $\sigma_z$ independent of the outcome $c^1$ since the unitary $V_E^{c^1}$ is cancelled. The procedure will be applied by each Eve$^m$ after his measurement before sending the post-measured state to the next Eve$^{m+1}$. If the system passed through only the unsharp measurements, the state received by the Bob at the end can be one of the $2^{n-1}$ potential states considering there are `n' Eves. This depend on all the Eves ($E^n$) outcomes (one for each combination $\overrightarrow{c}^{n-1} \equiv (c^1, c^2, \cdots, c^{n-1})$ of outcomes obtained by the Eves, these can be computed before the beginning of the QKD protocol). Any of these states can be written as:
\begin{equation}
	|\psi^n_{\overrightarrow{c}^{n-1}}\rangle = U_A^{\overrightarrow{c}^{n-1}} \otimes \mathbb{I}_2 \Big[ \cos(\theta_{\overrightarrow{c}^{n-1}}) |00\rangle + \sin(\theta_{\overrightarrow{c}^{n-1}}) |11\rangle \Big]
\end{equation}  
where the angles $\theta_{\overrightarrow{c}^{n-1}}$ and the matrix $U_A^{\overrightarrow{c}^{n-1}}$ both depend on the outcomes ${\overrightarrow{c}^{n-1}}$, on the initial angle $\theta_1$ and the sharpness parameters $\lambda_m$ of the Eves. Bob as usual performs the sharp projective measurements along $\sigma_z$ and $\sigma_x$ direction where as Alice performs measurements along $\sigma_z$ and her second measurement is 
\begin{equation}
	A_2^{\overrightarrow{c}^{n-1}} = U_A^{\overrightarrow{c}^{n-1}} \big[ \cos(\mu_{\overrightarrow{c}^{n-1}}) \sigma_z + \sin(\mu_{\overrightarrow{c}^{n-1}}) \sigma_x \big] (U_A^{\overrightarrow{c}^{n-1}})^{\dagger}
	\label{Alice_meas}
\end{equation}
where $\tan(\mu_{\overrightarrow{c}^{n-1}}) = \sin(2 \theta_{\overrightarrow{c}^{n-1}})$. These measurements maximally violate the fine-grained inequality and give maximal value of the secret key rate irrespective of the number of Eves. Note that the Alice's measurements depend on the outcome choices and unsharpness measurement parameters of the Eves and hence the probability of Alice performing measurement in the right direction decreases as the number of Eves increases. However, Alice will try to perform measurement in this direction because she along with Bob wants to maximally violate the inequality and achieve maximal secret key rate. This way unbounded number of Eves can also access the secret keys in this 1S-DIQKD network.


\section{Conclusions} \label{s4}
Quantum steerable correlations are the key resource for one-sided device-independent quantum key distribution task. The monogamy relations restrict the sharing of steerable resource, thus guaranteeing the security of the key distribution task. However, these monogamy restrictions are violated if the parties perform unsharp measurements \cite{Sasmal18}. It is fair to assume that adversary also has resources to do unsharp measurements and then it is legitimate to rethink about the security of such QKD networks.

In the present study, we address the security of 1S-DIQKD networks by considering the quantum unsharp measurement attacks by the eavesdroppers. We have considered a QKD network constituting two spatially separated spin-$\frac{1}{2}$ particles, one with Alice and the other one while going to Bob is intercepted by multiple eavesdroppers who gain some positive key rate by performing local quantum unsharp measurement attacks while preserving the quantum steerable correlations for the Bob. We find an upper bound on the number of such Eves.

We find that atmost 4, 3, and 2 number of Eves can access the secret keys when the secret key rate for each Eve is 0.1, 0.2 and 0.3 respectively. In each case, Bob has secret key rate atleast more than that Eve has access to. This is the unbiased case when both the measurements of Eves are unsharp. In ref. \cite{Sasmal18} at most two observers can share quantum steering when each observer is performing dichotomic measurements. However, we have obtained a maximum of 4 Eves when each party is performing dichotomic measurements showing an increment from the previous work. On the other hand in biased case when only the measurement corresponding to Eves' $\sigma_x$ measurement is unsharp and each Eve perform some unitary operation before sending the post-measured state to next Eve followed by the particular Alice's measurement (\ref{Alice_meas}), we get an unbounded number of Eves. Bob can assure the security of QKD protocol by having higher secret key rate and terminating the protocol in adverse scenario of drop in secret key rate below a threshold value. Determining this threshold as a function of Eve's key rate or environmental noise is worth looking in the future.

Before concluding, it may be noted that sequential sharing of bipartite steering \cite{Choi2020} have already been experimentally demonstrated. Our study raises a valid question towards the security of 1S-DIQKD networks that can be easily compromised using the available experimental techniques. Finally, the analysis presented here could be used to address the secrecy aspects of other important tasks like quantum secret sharing \cite{Hil} under quantum unsharp measurement attacks in near future.

\section{ ACKNOWLEDGEMENTS}
SG acknowledges S. N. Bose National Centre for Basic Sciences, Kolkata for the financial support. Authors acknowledge QWorld for organising QIntern 2021. SG thanks Shiladitya Mal for fruitful discussion.



\begin{thebibliography}{0}
\expandafter\ifx\csname natexlab\endcsname\relax\def\natexlab#1{#1}\fi
\expandafter\ifx\csname bibnamefont\endcsname\relax
  \def\bibnamefont#1{#1}\fi
\expandafter\ifx\csname bibfnamefont\endcsname\relax
  \def\bibfnamefont#1{#1}\fi
\expandafter\ifx\csname citenamefont\endcsname\relax
  \def\citenamefont#1{#1}\fi
\expandafter\ifx\csname url\endcsname\relax
  \def\url#1{\texttt{#1}}\fi
\expandafter\ifx\csname urlprefix\endcsname\relax\def\urlprefix{URL }\fi
\providecommand{\bibinfo}[2]{#2}
\providecommand{\eprint}[2][]{\url{#2}}

\end{thebibliography}


\begin{thebibliography}{1}

\bibitem{Gisin2002} Nicolas Gisin, Grégoire Ribordy, Wolfgang Tittel, and Hugo Zbinden, \emph{Quantum cryptography}, \href{https://link.aps.org/doi/10.1103/RevModPhys.74.145}{Rev. Mod. Phys. 74, 145 (2002).}

\bibitem{BB84} C. H. Bennett and G. Brassard, \emph{Quantum cryptography,public key distribution and coin tossing}, \href{https://researcher.watson.ibm.com/researcher/files/us-bennetc/BB84highest.pdf}{IEEE Press, New York, (1984).}

\bibitem{Bennet92} Charles H. Bennett, \emph{Quantum cryptography using any two nonorthogonal states}, \href{https://journals.aps.org/prl/abstract/10.1103/PhysRevLett.68.3121}{Phys. Rev. Lett. \textbf{68}, 3121 (1992).}

\bibitem{ekert} Artur K. Ekert, \emph{Quantum cryptography based on Bell's theorem
}, \href{https://journals.aps.org/prl/abstract/10.1103/PhysRevLett.67.661}{Phys. Rev. Lett. 67, 661  (1991)}

\bibitem{scarani2009} V. Scarani, H. Bechmann-Pasquinucci, N. J. Cerf, M. Dušek, N. Lütkenhaus, and M. Peev, \emph{The security of practical quantum key distribution
}, \href{https://journals.aps.org/rmp/abstract/10.1103/RevModPhys.81.1301}{Rev. Mod. Phys. 81, 1301 (2009)}


\bibitem{Mayers97} Dominic Mayers, \emph{Unconditionally Secure Quantum Bit Commitment is Impossible}, \href{https://journals.aps.org/prl/abstract/10.1103/PhysRevLett.78.3414}{Phys. Rev. Lett. 78, 3414 (1997)}

\bibitem{Mayers2001} Dominic Mayers, \emph{Unconditional security in quantum cryptography}, \href{https://dl.acm.org/doi/10.1145/382780.382781}{J.ACM 48, 351 (2001)}

\bibitem{Deutsch96} David Deutsch, Artur Ekert, Richard Jozsa, Chiara Macchiavello, Sandu Popescu, and Anna Sanpera, \emph{Quantum Privacy Amplification and the Security of Quantum Cryptography over Noisy Channels}, \href{https://journals.aps.org/prl/abstract/10.1103/PhysRevLett.77.2818}{Phys. Rev. Lett. 77, 2818 (1996)}

\bibitem{Lo99} Hoi-Kwong Lo and H. F. Chau, \emph{Is Quantum Bit Commitment Really Possible?}, \href{https://journals.aps.org/prl/abstract/10.1103/PhysRevLett.78.3410}{Phys. Rev. Lett. 78, 3410 (1997)}

\bibitem{Shor2000} Peter W. Shor and John Preskill
, \emph{Simple Proof of Security of the BB84 Quantum Key Distribution Protocol}, \href{https://journals.aps.org/prl/abstract/10.1103/PhysRevLett.85.441}{Phys. Rev. Lett. 85, 441 (2000)}

\bibitem{Brassard2000} Gilles Brassard, Norbert Lütkenhaus, Tal Mor, and Barry C. Sanders
, \emph{Limitations on Practical Quantum Cryptography}, \href{https://journals.aps.org/prl/abstract/10.1103/PhysRevLett.85.1330}{Phys. Rev. Lett. 85, 1330 (2000)}

\bibitem{Norbert2000} Norbert Lütkenhaus, \emph{Security against individual attacks for realistic quantum key distribution}, \href{https://journals.aps.org/pra/abstract/10.1103/PhysRevA.61.052304}{Phys. Rev. A. 61, 052304 (2000)}

\bibitem{Branciard12} Cyril Branciard, Eric G. Cavalcanti, Stephen P. Walborn, Valerio Scarani, and Howard M. Wiseman, \emph{One-sided device-independent quantum key distribution: Security, feasibility, and the connection with steering}, \href{https://journals.aps.org/pra/abstract/10.1103/PhysRevA.85.010301}{Phys. Rev. A. 85, 010301(R) (2012)}

\bibitem{steerreview} R. Uola, A. C. S. Costa, H. C. Nguyen, and O. Guhne, \emph{Quantum steering}, \href{https://journals.aps.org/rmp/abstract/10.1103/RevModPhys.92.015001}{Rev. Mod. Phys. {\bf 92}, 015001 (2020)}.
\bibitem{Schr1} E. Schr\"{o}dinger, \emph{Discussion of Probability Relations between Separated Systems}, \href{https://www.cambridge.org/core/journals/mathematical-proceedings-of-the-cambridge-philosophical-society/article/discussion-of-probability-relations-between-separated-systems/C1C71E1AA5BA56EBE6588AAACB9A222D}{ Proc. Cambridge Philos. Soc. {\bf 31}, 553 (1935).}
\bibitem{Schr2} E. Schr\"{o}dinger, \emph{Probability relations between separatedsystems}, \href{https://www.cambridge.org/core/journals/mathematical-proceedings-of-the-cambridge-philosophical-society/article/probability-relations-between-separated-systems/641DDDED6FB033A1B190B458E0D02F22}{ Proc. Cambridge Philos. Soc. {\bf 32}, 446 (1936).}

\bibitem{Wise1} H. M. Wiseman, S. J. Jones, and A. C. Doherty, \emph{Steering, Entanglement, Nonlocality, and the Einstein-PodolskyRosen Paradox}, \href{https://journals.aps.org/prl/abstract/10.1103/PhysRevLett.98.140402}{Phys. Rev. Lett. {\bf 98}, 140402 (2007).}
\bibitem{Wise2} S. J. Jones, H. M. Wiseman, and A. C. Doherty, \emph{Entanglement, Einstein-Podolsky-Rosen correlations, Bell nonlocality, and steering}, \href{https://journals.aps.org/pra/abstract/10.1103/PhysRevA.76.052116}{Phys. Rev. A. {\bf 76}, 052116 (2007).}
\bibitem{Cavalcanti1} E. G. Cavalcanti, S. J. Jones, H. M. Wiseman and M. D. Reid, \emph{Experimental criteria for steering and the Einstein-Podolsky-Rosen paradox}, \href{https://journals.aps.org/pra/abstract/10.1103/PhysRevA.80.032112}{Phys. Rev. A {\bf 80}, 032112 (2009).} 
\bibitem{Cavalcanti2} J. Schneeloch, C. J. Broadbent, S. P. Walborn, E. G. Cavalcanti, and J. C. Howell, \emph{Einstein-Podolsky-Rosen steering inequalities from entropic uncertainty relations}, \href{https://journals.aps.org/pra/abstract/10.1103/PhysRevA.87.062103}{Phys. Rev. A {\bf 87}, 062103 (2013).} 
\bibitem{SGupta21} Shashank Gupta, Debarshi Das, and A. S. Majumdar, \emph{Distillation of genuine tripartite Einstein-Podolsky-Rosen steering
}, \href{https://journals.aps.org/pra/abstract/10.1103/PhysRevA.104.022409}{Phys. Rev. A {\bf 104}, 022409 (2021).}
\bibitem{epr} A. Einstein, B. Podolsky, and N. Rosen,  \emph{Can Quantum-Mechanical Description of Physical Reality Be Considered Complete?}, \href{https://journals.aps.org/pr/abstract/10.1103/PhysRev.47.777}{Phys. Rev. {\bf 47}, 777 (1935).}

\bibitem{Pramanik} T. Pramanik, M. Kaplan and A. S. Majumdar, \emph{Fine-grained Einstein-Podolsky-Rosen steering inequalities}, \href{https://journals.aps.org/pra/abstract/10.1103/PhysRevA.90.050305}{Phys. Rev. A {\bf 90}, 050305(R) (2014).} 

\bibitem{Sasmal18} Souradeep Sasmal, Debarshi Das, Shiladitya Mal, and A. S. Majumdar, \emph{Steering a single system sequentially by multiple observers
}, \href{https://journals.aps.org/pra/abstract/10.1103/PhysRevA.98.012305}{Phys. Rev. A {\bf 98}, 012305 (2018).}

\bibitem{Gupta21} Shashank Gupta, Ananda G. Maity, Debarshi Das, Arup Roy, and A. S. Majumdar, \emph{Genuine Einstein-Podolsky-Rosen steering of three-qubit states by multiple sequential observers}, \href{https://journals.aps.org/pra/abstract/10.1103/PhysRevA.103.022421}{Phys. Rev. A {\bf 103}, 022421 (2021).}



\bibitem{Curchod17} F. J. Curchod, M. Johansson, R. Augusiak, M. J. Hoban, P. Wittek, and A. Acín, \emph{Unbounded randomness certification using sequences of measurements}, \href{https://journals.aps.org/pra/abstract/10.1103/PhysRevA.95.020102}{Phys. Rev. A {\bf 95}, 020102(R) (2017).}

\bibitem{sygp} R. Silva, N. Gisin, Y. Guryanova, and Sandu Popescu, \emph{Multiple Observers Can Share the Nonlocality of Half of an Entangled Pair by Using Optimal Weak Measurements}, \href{https://journals.aps.org/prl/abstract/10.1103/PhysRevLett.114.250401}{Phys. Rev. Lett. {\bf 114}, 250401 (2015).}
\bibitem{majumdar} S. Mal, A. S. Majumdar, D. Home, \emph{Sharing of Nonlocality of a Single Member of an Entangled Pair of Qubits Is Not Possible by More than Two Unbiased Observers on the Other Wing}, \href{http://www.mdpi.com/2227-7390/4/3/48/htm}{Mathematics {\bf 4} 48 (2016).}
\bibitem{pb1} P. Busch, \emph{Unsharp reality and joint measurements for spin observables},  \href{https://journals.aps.org/prd/abstract/10.1103/PhysRevD.33.2253}{Phys. Rev. D {\bf 33}, 2253 (1986).}
\bibitem{pb2} P. Busch, M. Grabowski, and P. J. Lahti, \emph{Operational Quantum Physics} (Springer-Verlag, Berlin, 1997).
\bibitem{umn1} M. Banik, Md. R. Gazi, S. Ghosh, and G. Kar, \emph{Degree of complementarity determines the nonlocality in quantum mechanics}, \href{https://journals.aps.org/pra/abstract/10.1103/PhysRevA.87.052125}{Phys. Rev. A {\bf 87}, 052125 (2013).}
\bibitem{umn2} P. Busch, T. Heinosaari, J. Schultz, and N. Stevens, \emph{Comparing the degrees of incompatibility inherent in probabilistic physical theories}, \href{https://iopscience.iop.org/article/10.1209/0295-5075/103/10002}{Europhys. Lett. {\bf 103}, 10002 (2013).}
\bibitem{une1} S. K. Choudhary, T. Konrad, and H. Uys, \emph{Implementation schemes for unsharp measurements with trapped ions}, \href{https://journals.aps.org/pra/abstract/10.1103/PhysRevA.87.012131}{Phys. Rev. A {\bf 87}, 012131 (2013).}

\bibitem{exp1} M.-J. Hu, Z.-Y. Zhou, X.-M. Hu, C.-F. Li, G.-C. Guo, and Y.-S. Zhang, \emph{Observation of non-locality sharing among three observers with one entangled pair via optimal weak measurement}, \href{https://www.nature.com/articles/s41534-018-0115-x}{npj Quantum Information {\bf 4}, 63 (2018).}
\bibitem{exp2} M. Schiavon, L. Calderaro, M. Pittaluga, G. Vallone, and P. Villoresi, \emph{Three-observer Bell inequality violation on a two-qubit entangled state}, \href{http://iopscience.iop.org/article/10.1088/2058-9565/aa62be/meta}{Quantum Sci. Technol. \textbf{2} 015010 (2017).}

\bibitem{Choi2020} Y.-H. Choi, S. Hong, T. Pramanik, H.-T. Lim, Y.-S. Kim, H. Jung, S.-W. Han, S. Moon, and Y.-W. Cho, \emph{Demonstration of simultaneous quantum steering by multiple observers via sequential weak measurements}, \href{https://www.osapublishing.org/optica/fulltext.cfm?uri=optica-7-6-675&id=432421}{Optica {\bf 7}, 675 (2020).}
\bibitem{exp4} G. Foletto, L. Calderaro, G. Vallone, and P. Villoresi, \emph{Experimental demonstration of sequential quantum random access codes}, \href{https://journals.aps.org/prresearch/abstract/10.1103/PhysRevResearch.2.033205}{Phys. Rev. Research {\bf 2}, 033205 (2020).}
\bibitem{rac1} H. Anwer, S. Muhammad, W. Cherifi, N. Miklin, A. Tavakoli, and M. Bourennane, \emph{Experimental Characterization of Unsharp Qubit Observables and Sequential Measurement Incompatibility via Quantum Random Access Codes}, \href{https://journals.aps.org/prl/abstract/10.1103/PhysRevLett.125.080403}{Phys. Rev. Lett. {\bf 125}, 080403 (2020).}

\bibitem{Cavalcanti15} Eric G. Cavalcanti, Christopher J. Foster, Maria Fuwa, and Howard M. Wiseman, \emph{Analog of the Clauser-Horne-Shimony-Holt inequality for steering
}, \href{https://www.osapublishing.org/josab/abstract.cfm?uri=josab-32-4-A74}{J. Opt. Soc. Am. B 32, A74 (2015)}

\bibitem{Hil} M. Hillery, V. Buzek, and A. Berthiaume, \emph{Quantum secret sharing}, \href{https://journals.aps.org/prl/abstract/10.1103/PhysRevLett.68.3121}{Phys. Rev. Lett. 68, 3121 (1992)}
























\end{thebibliography}
\end{document}